\documentclass[aps,pre,reprint,superscriptaddress]{revtex4-1}
\usepackage{amsmath}
\usepackage{amssymb}
\usepackage{graphicx}
\usepackage{dcolumn}
\usepackage{bm}
\usepackage[utf8]{inputenc}
\usepackage[T1]{fontenc}
\usepackage{mathptmx}
\usepackage{xcolor}
\usepackage{multirow}

\begin{document}


\title{A generating function approach to the growth rate of random matrix products}


\author{Naranmandula Bao}
\affiliation{College of Physics and Electronics Information, Inner Mongolian University for Nationalities, Inner Mongolia, China 028000}
\author{Junbiao Lu}
\affiliation{School of Science, Beijing University of Posts and Telecommunications, Beijing, China 100876}
\author{Yueheng Lan}
\email[Corresponding author:]{lanyh@bupt.edu.cn}
\affiliation{School of Science, Beijing University of Posts and Telecommunications, Beijing, China 100876}
\affiliation{State Key Lab of Information Photonics and Optical Communications, Beijing University of Posts and Telecommunications, Beijing, China 100876}
\date{\today}
\begin{abstract}
Random matrix products arise in many science and engineering problems. An efficient evaluation of its growth rate is of great interest to researchers in diverse fields. In the current paper, we reformulate this problem with a generating function approach, based on which two analytic methods are proposed to compute the growth rate. The new formalism is demonstrated in a series of examples including an Ising model subject to on-site random magnetic fields, which seems very efficient and easy to implement. Through an extensive comparison with numerical computation, we see that the analytic results are valid in a regime of considerable size.
\end{abstract}
\maketitle
\section{INTRODUCTION}
\label{introduction}
Evolution of linear systems could be described by repeated production of state vectors with the corresponding matrices characterizing linear dynamics. In the presence of randomness in the dynamics, the evolution may be analyzed with random matrix products. As a result, in diverse fields of science and engineering~ \cite{cris93prod,cris94prod,emb99tre,sch57dis,bou85sch,dia86imag,tul86app}, random matrix products show up ubiquitously and has been a hot subject for intensive study for a long time~\cite{cris93prod}. In certain cases, the investigation
is extended to the study of random products of operators as a generalization to infinite-dimensions~\cite{bell54oper,fur63prod,dia99iter}.\\
Among different aspects of the random matrix products, the computation of the growth rate of their norms is of fundamental importance~\cite{cris93prod,sir01rand,osh02reac}, which is related to the spectra of random oscillator chain~\cite{dys53lin,sch57dis}, the density of states in a disordered quantum system~\cite{bou85sch,der00sch} or the famous Anderson localization~\cite{com023and,der84and} and its application in quantum mechanics~\cite{gold98dis,fein99loc,hat96loc}. However, few cases exist which can be solved with an exact analytic computation~\cite{sir01rand}. The most common practice remains to be Monte carlo simulation which is easy to implement and thus applied widely~\cite{cris93prod,emb99tre,vis99,tre01comp,van01gen}. It is undoubtedly a good way to get a quick estimation of the growth rate. However, as well known, the convergence of the Monte carlo computation is slow, which in many cases prevents an accurate evaluation, especially when the growth rate is small. Moreover, it is hard to extract analytic dependence on parameters in a pure numerical computation. For good accuracy, sometimes cycle expansions could be used to extract the growth rate but its implementation quickly becomes rather involving if the number of the sampled matrices is large~\cite{main92cyc,bai07cyc}. The analytic development concentrates on various perturbation methods. Weak or strong order expansion is invented in the case that there is a small or large parameter~\cite{der87wea,ben06weak,sir01rand}. Nevertheless, most of their application is focused on low-dimensional matrices and the computation sometimes turns out very technical even for low-order expansions~\cite{sir01rand}. \\
In a recent paper~\cite{lan11}, we proposed a generating function approach for the evaluation of growth rates of random Fibonacci sequences, which is very efficient and good for both analytic or numerical computation. Later on, the method is extended to random sequences with multi-step memory and a conjectured general structure in the generating function iteration is proved~\cite{lan13}. As a result, all the earlier computations carry through in the generalized case. As the random sequence is
a special case of random matrix products~\cite{cris93prod}, it is thus tempting to see if this new formalism could be further extended to an even more general case. After some endeavour, we succeeded in deriving such a new formulation. In the current paper, we present our generating function approach to the evaluation of growth rates of random matrix products. In a variety of examples given below, analytic expansions or even exact expressions in certain cases for growth rates are easily derived, for matrices with various dimensions or with different distributions.  \\
The paper is organized as follows. In section \ref{sec:form}, a general formulation of the random matrix products is given based on the generating function approach. In the next three sections, two different methods are proposed for the evaluation of growth rates based on the new formulation. The first one is a direct method which is easy to implement and discussed in section \ref{sec:dir}. The other one is an approach based on invariant polynomials which is more general and discussed in section \ref{sec:pol}. In these three sections, analytic and numerical results are compared to check the validity of the new formulation. Then in section \ref{sec:isi}, the methods are applied to the calculation of the free energy of a random Ising model. Finally, problems and possible future directions are discussed in section \ref{sec:dis}.
\section{Formulation based on the generating function}
\label{sec:form}
Here, after a brief review of the generating function technique in the evaluation of the grow rate of a random sequence,
we use $2\times 2$
random matrices to give a taste of the new method. Then, a very general formulation is presented, which is good for random matrices of any dimension or with any sampling rate.
\subsection{A brief review of the generating function approach to stochastic sequences}
\label{sec:fibo}
In a previous paper~\cite{lan11}, we used a generating function approach to treat the generalized random Fibonacci sequence
\begin{equation}
x_{n+1}=x_n\pm \beta x_{n-1}
\,,\label{eq:ran-fib}
\end{equation}
where the plus or the minus sign is picked up with equal probability, and $\beta>0$ is a parameter controlling the size of the random term. In most cases, $|x_n|\sim \exp(\lambda n)$ for $n \to \infty$,
where $\lambda$ is the growth rate of the random sequence, which could be positive indicating a growth or negative indicating a shrinkage of the magnitude of a typical term in the sequence in the asymptotic limit. The zero value of $\lambda$ usually signifies a phase transition~\cite{tre01comp}. In \cite{lan11}, we proposed a generating function which calculates $\lambda$ as
\begin{equation}
\lambda=\frac{1}{2}\lim_{n\to\infty}g_n(0)
\,,\label{eq:lam-fib}
\end{equation}
where $\{g_n(x)\}_{n=0,1,2,\cdots}$ is a sequence of generating functions which satisfy the recursion
relation
\begin{equation}
g_{n+1}(x)=\frac{1}{2}\left[g_n(\frac{\beta}{1+x})+g_n(-\frac{\beta}{1+x})\right]
\,,\label{eq:recur-fib}
\end{equation}
with $g_0(x)=\ln (1+x)^2$. The actual evaluation of the growth rate could be done either numerically or analytically by virtue of Eq.~(\ref{eq:recur-fib}). One pleasant surprise is that for $\beta$ small a direct expansion in $\beta$ can be obtained trivially by a Taylor expansion of $g_n(0)$ with small $n$
\begin{equation}
\lambda(\beta)=-\frac{1}{2}\beta^2-\frac{7}{4}\beta^4-\frac{29}{3}\beta^6+O(\beta^8)
\,.\label{eq:exp-fib}
\end{equation}
For example, the expansion~(\ref{eq:exp-fib}) is obtained with $n=3$.
Interestingly, the above formulation could be extended to a stochastic sequence with $n$-step memory~\cite{lan13}. Here is an example with $3-$step memory
\begin{equation}
x_{n+1}=x_n\pm\beta x_{n-1}\pm\gamma x_{n-2}
\,,\label{eq:ran-seq3}
\end{equation}
where $\beta,\gamma>0$ and the plus or minus sign is taken with equal probability. Similar expressions for the growth rate and the recursion relation are obtained, which enables a convenient derivation of an
asymptotic expansion of the growth rate as well for small $\beta$ and $\gamma$~\cite{lan13}.
As we know, a stochastic sequence problem could be formulated in the form of random matrix products~\cite{cris93prod,tre01comp,main92cyc}. For example, the generalized random Fibonacci sequence could be written as
\begin{equation}
Y_{n+1}=A_n Y_n
\,,\label{eq:mat-fib}
\end{equation}
where $Y_n=(x_{n}\,,x_{n-1})^t$ is a $2-$d vector and
\begin{equation}
A_n=\left(\begin{array}{cc}1 & \beta \\
                           1 & 0  \end{array}\right) \mbox{ or }
    \left(\begin{array}{cc}1 & -\beta \\
                           1 & 0  \end{array}\right)
\,,\label{eq:a-fib}
\end{equation}
samples two possible matrices with equal probability in the iteration. However, generally, it is not possible to write a random matrix product as a stochastic sequence. One natural question is whether the above formalism for computing the growth rate could be further extended to a general matrix. How about random matrices with more involved distributions? We have positive answers to all these questions as shown below.
\subsection{Generating function for 2-d matrix products}
To illustrate the computation, we first consider a special case: a 2-d matrix product with a simple distribution. The random matrix $A_n$ in Eq.~(\ref{eq:a-fib}) is
a special case of the matrix below
\begin{equation}
A_{n\pm}=\left(\begin{array}{cc}a & b \\
                           c & d  \end{array}\right) \pm \beta
    \left(\begin{array}{cc}e & r \\
                           g & h  \end{array}\right)
\,,\label{eq:2d-fib}
\end{equation}
where the plus sign is taken with probability $p$ and the minus sign with $1-p$. The matrix before the plus minus sign is the major part of the random matrices since both matrices reduce to it when $\beta=0$. Similar to the case of a stochastic sequence, a family of generating functions could be defined by the recursion relation
\begin{eqnarray}\label{eq:2d-recur}
\nonumber
f_{n+1}(x)=&&p f_n\left(\frac{b+dx+\beta(r+hx)}{a+cx+\beta(e+gx)}\right)\\
           &&+(1-p) f_n\left(\frac{b+dx-\beta(r+hx)}{a+cx-\beta(e+gx)}\right),
\end{eqnarray}
with the starting function
\begin{eqnarray}\label{eq:2d-init}
\nonumber
f_0(x)=&&p\ln[a+cx+\beta(e+gx)]\\
       &&+(1-p)\ln[a+cx-\beta(e+gx)].
\end{eqnarray}
The maximum Lyapunov exponent (MLE) is then $\lambda={\lim_{n\to\infty}}f_n(0)$. Note that when the parameters take appropriate values
in Eq.~(\ref{eq:2d-fib}) (a=c=r=1\,,b=d=e=g=h=0\,,p=1/2) we recover all the equations for the random
Fibonacci sequence.
It is easy to see why the functions defined by Eq.~(\ref{eq:2d-recur}) and Eq.~(\ref{eq:2d-init}) are good
for computing the MLE. In the argument of the logarithm in Eq.~(\ref{eq:2d-init}), the constant and the coefficient of $x$ could be viewed as the first and second component of a $2-$d vector $Y_1$ (see Eq.~(\ref{eq:mat-fib})). The two logarithms in Eq.~(\ref{eq:2d-init}) encodes the random vector $Y_1=A_0 Y_0$, where $A_0$ is taken from Eq.~(\ref{eq:2d-fib})) and $Y_0=(1\,,0)^t$. Thus we obtain
$Y_1=(a+\beta e\,,c+\beta g)^t$ with probability $p$ and $Y_1=(a-\beta e\,,c-\beta g)^t$ with probability $1-p$.
Similarly, all the $Y_n$'s are encoded in the logarithmic terms in $f_{n-1}(x)$ and the iteration Eq.~(\ref{eq:2d-recur}) corresponds to the matrix multiplication Eq.~(\ref{eq:mat-fib}), which produces all the possible $Y_{n+1}$'s, being embedded in $f_{n}(x)$. In each logarithmic term of $f_{n}(x)$, the argument is in fact a ratio of linear functions of the components of $Y_{n+1}$ and its preimage $Y_n$. The coefficient before the logarithm is the probability for this ratio to occur.
\begin{figure*}
\begin{eqnarray}\label{eq:2d-n2}
\nonumber
f_1(x)=&&p^2 \ln\left(\frac{(a+\beta e)^2+(c+\beta g)(b+\beta r)+x((a+\beta e)(c+\beta g)+(c+\beta g)(a+\beta h))}{a+cx+\beta(e+gx)}\right)\\
\nonumber
&&+p(1-p) \ln\left(\frac{(a-\beta e)(a+\beta e)+(c-\beta g)(b+\beta r)+x((a-\beta e)(c+\beta g)+(c-\beta g)(a+\beta h))}{a+cx+\beta(e+gx)}\right)\\
\nonumber
&&+(1-p)^2 \ln\left(\frac{(a-\beta e)^2+(c-\beta g)(b-\beta r)+x((a-\beta e)(c-\beta g)+(c-\beta g)(a-\beta h))}{a+cx-\beta(e+gx)}\right)\\
&&+(1-p)p \ln\left(\frac{(a+\beta e)(a-\beta e)+(c+\beta g)(b-\beta r)+x((a+\beta e)(c-\beta g)+(c+\beta g)(a-\beta h))}{a+cx-\beta(e+gx)}\right).
\end{eqnarray}
\end{figure*}

For example, in the first logrithmic term of Eq.~(\ref{eq:2d-n2}), the argument is a ratio of two linear functions in $x$. In its denominator, we see the components of $Y_1$ and those of $Y_2$ in the numerator. Four possible values for $Y_2$ are embedded in the four terms of $f_1(x)$ with the probabilities $p^2\,,p(1-p)\,,(1-p)^2\,,(1-p)p$. Depending on the value of $x$, the argument gives different ratios and the function $f_1(x)$ is a weighted sum of logrithms of these ratios. If we take $x=0$, the argument is the ratio of the first components of consecutive random vectors; if $x=1$ is taken, the argument is the ratio of the sums of the two components. In general, the arguments of the $2^{n+1}$ logarithms in $f_n(x)$ give all possible such ratios. With $n\to \infty$, the magnitudes of these ratios will approach a stationary distribution and the coefficients before the logarithm are the corresponding probabilities. Thus, $f_n(0)$ appoaches the MLE~\cite{vis99}. From the argument above, it can
also be seen that in the limit $n\to \infty$,
the function $f_n(x)$ will approach a constant - the MLE, almost everywhere.
\subsection{Generating function for general matrix products}
The above argument could be easily extended to the general case: the set of $(k+1) \times (k+1)$ matrices $B(\alpha)$ with arbitrary dimension and sampling rates $P(\alpha)$. As before, each $(k+1)-$dim vector $\mathbf{a}=(a_1\,,a_2\,,\cdots\,,a_{k+1})^t$ could be encoded by a linear expression
\begin{equation}
\ell(\mathbf{a})=a_1+a_2 z_2+\cdots+a_{k+1} z_{k+1}
\,,\label{eq:n-linear}
\end{equation}
where $z_2\,,\cdots\,,z_{k+1}$ are formal variables to encode the components of the vector. To encode the random products, each matrix $B(\alpha)$
corresponds to a transformation $T(\alpha)$ defined as
\begin{equation}
T(\alpha)\circ f(z_2\,,z_3\,,\cdots\,,z_{k+1})=f(\tilde{z}_2\,,\tilde{z}_3\,,\cdots\,,\tilde{z}_{k+1})
\,,\label{eq:n-T}
\end{equation}
where
\begin{equation}
\tilde{z}_i=\frac{\sum_{j=1}^{k+1} B(\alpha)_{ji}z_j}{\sum_{j=1}^{k+1} B(\alpha)_{j1} z_j}\,,i=2\,,3\,,\cdots
\,,\label{eq:n-zi}
\end{equation}
with the convention $z_1=1\,,\tilde{z}_1=1$. In this notation, the recursion relation could be written as
\begin{equation}
f_{n+1}(z_2\,,z_3\,,\cdots\,,z_{k+1})=\int d\alpha P(\alpha) T(\alpha)\circ f_n(z_2\,,z_3\,,\cdots\,,z_{k+1})
\,,\label{eq:n-recur}
\end{equation}
where again $P(\alpha)$ denotes the sampling rate of the matrix $B(\alpha)$. The starting function is
\begin{equation}
f_{0}(z_2\,,z_3\,,\cdots\,,z_{k+1})=\int d\alpha P(\alpha) \ln (\sum_j B(\alpha)_{j1} z_j)
\,,\label{eq:n-start}
\end{equation}
and as before the MLE could be obtained as
\begin{equation}
\lambda=\lim_{n\to \infty} f_n(0)
\,.\label{eq:n-lam}
\end{equation}
If the distribution is discrete, the integration in Eq.~(\ref{eq:n-recur}) should be replaced by a summation.
Note that Eq.~(\ref{eq:n-lam}) is exact and the problem that remains is the evaluation of $f_n(\mathbf{z})$ in the limit $n\to \infty$. As explained in \cite{lan11,lan13}, the recursion relation could be used directly in a numerical computation. In the presence of small parameters, various expansion techniques could be developed. Here, we will be concentrating on two of them as demonstrated below with different levels of sophistication.
\section{Direct expansion}
\label{sec:dir}
In section \ref{sec:fibo}, we see that the expansion in $\beta$ is easily carried out to high orders just by several iterations of the recursion relation. This crisp feature could be maintained if the matrix $A$ has a special structure. We will give several examples in the following to illustrate this point. To see the validity of the derived analytic expressions, Monte Carlo simulation is used for comparison. Each MLE is generated by carrying out one million Monte Carlo steps.
\subsection{Expansion for 2-d matrices}
\begin{figure*}
\begin{minipage}{0.48\linewidth}
\centerline{\includegraphics[width=8.5cm]{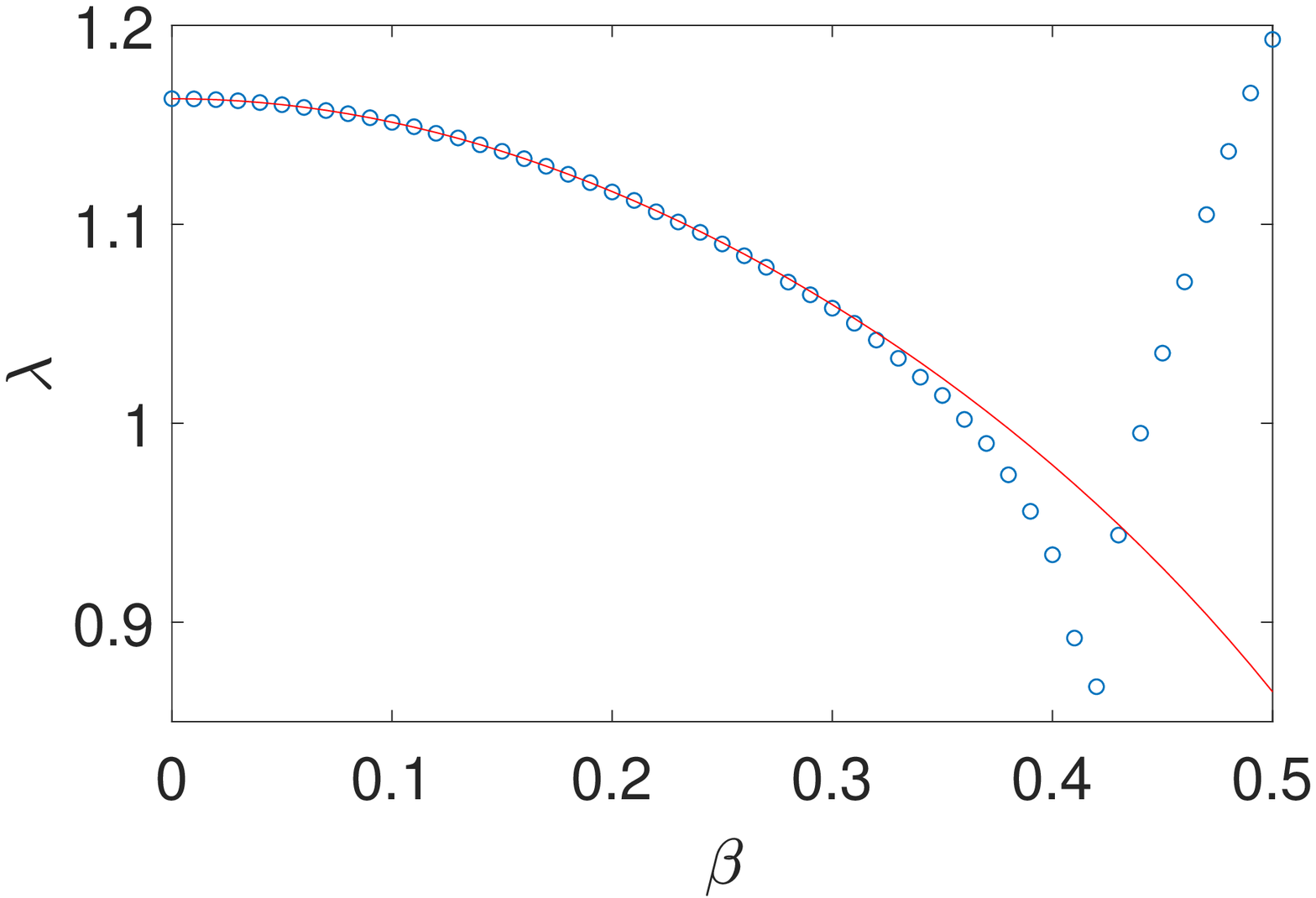}}
\centerline{(a)}
\end{minipage}
\hfill
\begin{minipage}{0.48\linewidth}
\centerline{\includegraphics[width=8.5cm]{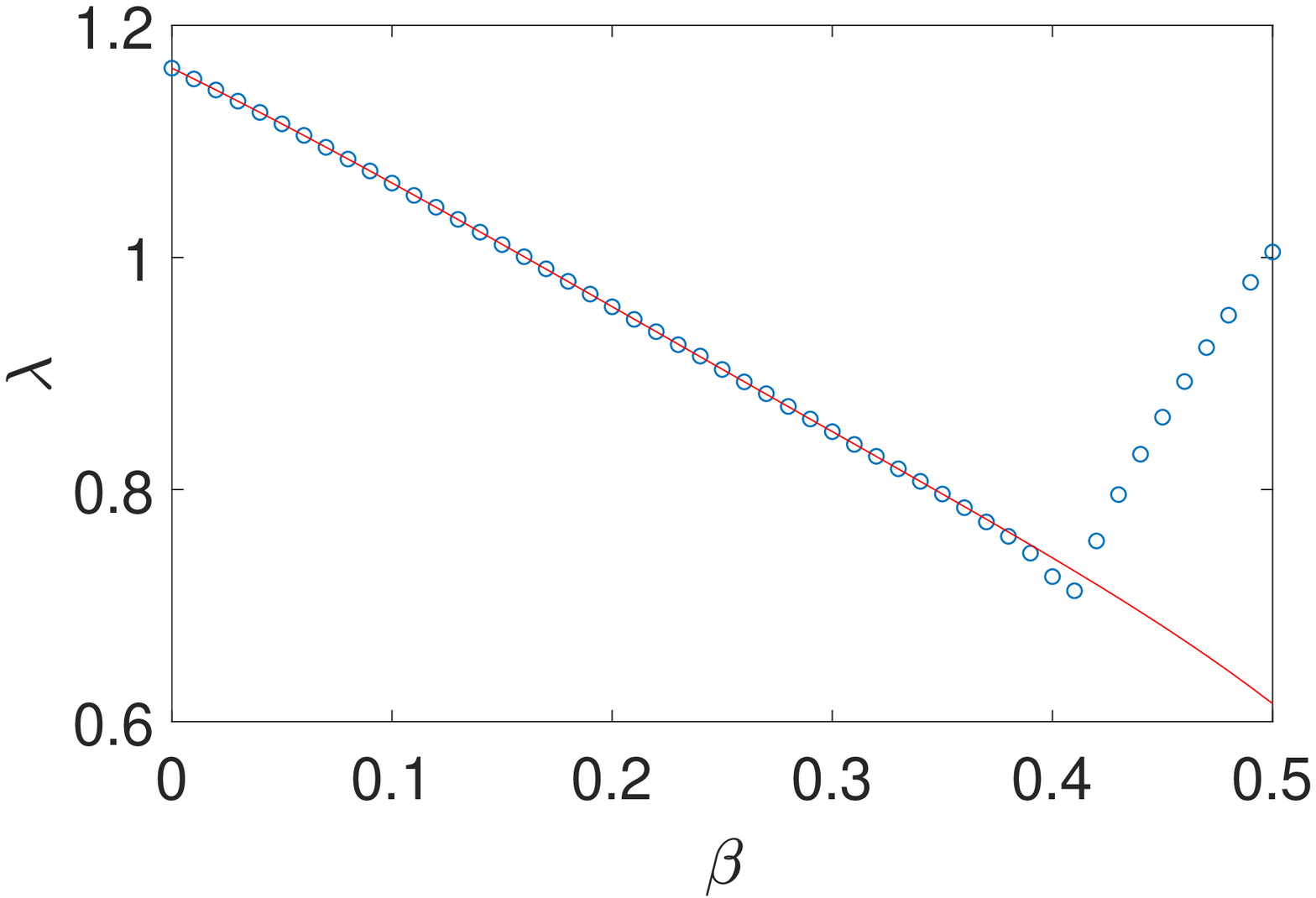}}
\centerline{(b)}
\end{minipage}
\vfill
\begin{minipage}{0.48\linewidth}
\centerline{\includegraphics[width=8.5cm]{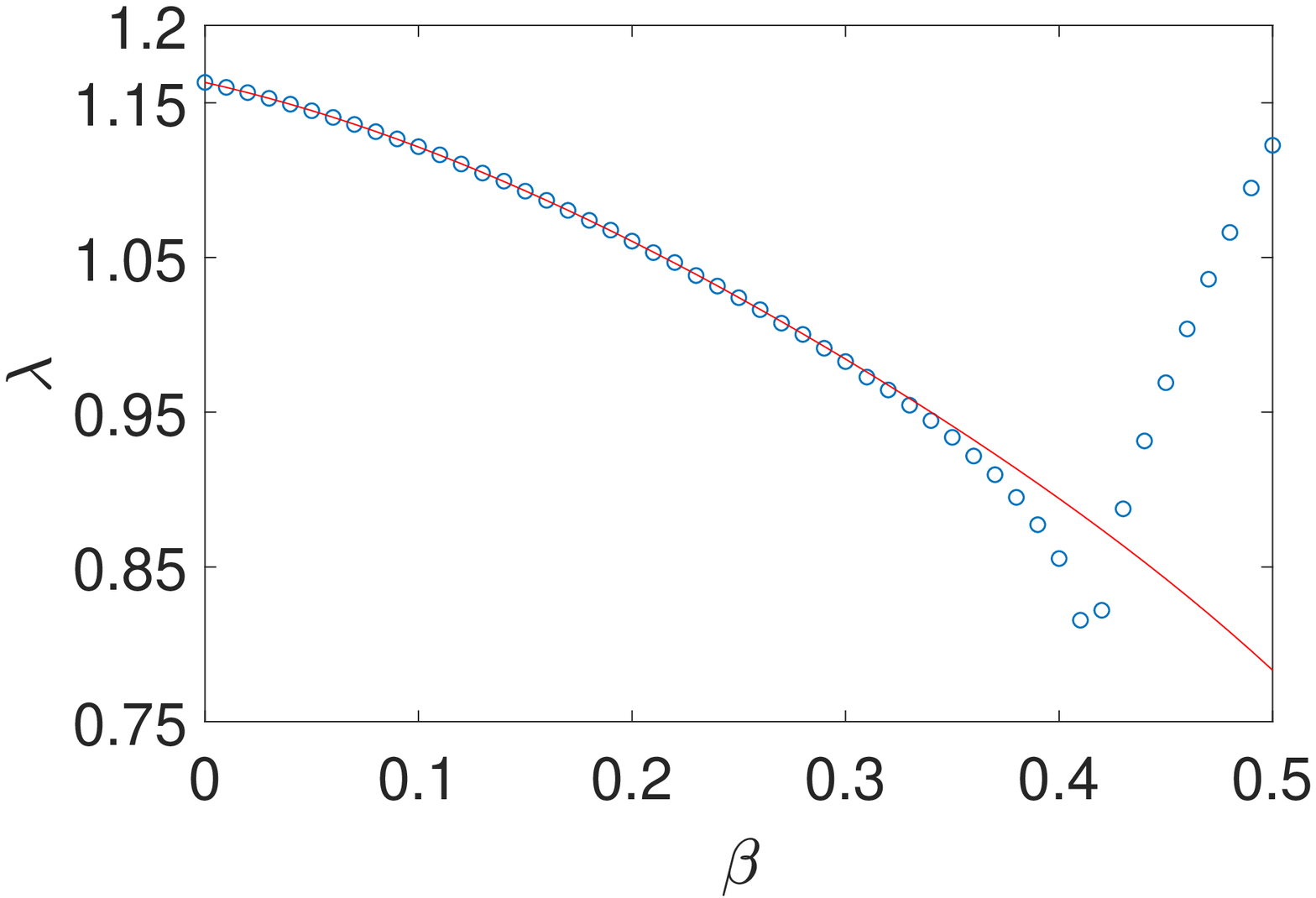}}
\centerline{(c)}
\end{minipage}
\hfill
\begin{minipage}{0.48\linewidth}
\centerline{\includegraphics[width=8.5cm]{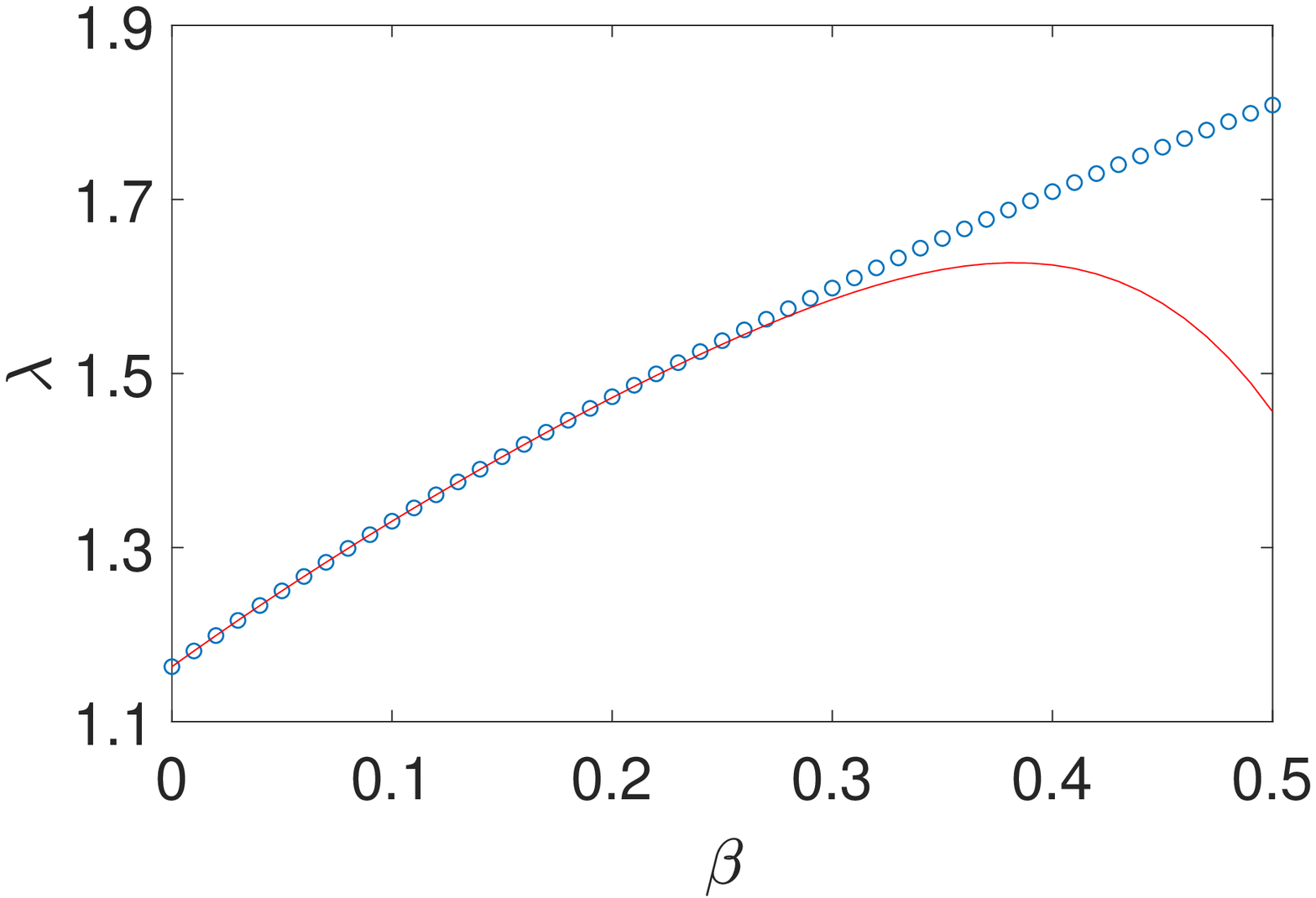}}
\centerline{(d)}
\end{minipage}
\caption{(Color online) The dependence of the MLE on
$\beta$ for different probabilities $p$ of taking the plus and minus sign in Eq.~\ref{eq:2d-dir1}.
(a)$p=0.5\,,0.5$;(b) $p=0.2\,,0.8$;(c) $p=0.4\,,0.6$;(d) the result for the system in Eq.~(\ref{eq:2dp-b}). The Monte carlo results are marked with (blue) circles while the analytic approximation is plotted with (red) solid lines.
} \label{f:dir-2dlam}
\end{figure*}

To verify the validity of our procedure, we use the $2\times 2$ matrices below
\begin{equation}
A=\left(\begin{array}{cc}a & b \\
                           c & d  \end{array}\right) \pm \beta
    \left(\begin{array}{cc}e & r \\
                           g & h  \end{array}\right)
\,.\label{eq:2d-dir}
\end{equation}
In the above equation, if $b=\mu a\,, d=\mu c$, the argument substitution in Eq.~(\ref{eq:2d-recur}) becomes $x\to \mu \pm \beta R(x) $, where $R(x)$ is a rational function of $x$. Thus, each additional iteration gives a higher order expression of $\lambda$ in $\beta$ when $\beta$ is small. \\
More explicitly, if we take
\begin{equation}
A=\left(\begin{array}{cc}  3 & 0.3 \\
                           2 & 0.2  \end{array}\right) \pm \beta
    \left(\begin{array}{cc}3 & 2 \\
                           6 & 5  \end{array}\right)
\,,\label{eq:2d-dir1}
\end{equation}
then a few iterations result in
\begin{eqnarray}\label{eq:dir-2dlam1}
\nonumber
\lambda(\beta)=&&4\ln(2)-\ln(5)-\frac{156025}{131072}\beta^2+\frac{12034995625}{17179869184}\beta^4 \\
               &&-\frac{2397118282953125}{844424930131968}\beta^6+O(\beta^8).
\end{eqnarray}
In Fig.~\ref{f:dir-2dlam}(a), the expansion Eq.~(\ref{eq:dir-2dlam1}) is compared with the results from the Monte carlo simulation. When $\beta<0.3$, the two agree very well while the error rises quickly after
$\beta=0.35$. Especially, the valley at $\beta \sim 0.42$ is not present in the analytic result. This is because the expression Eq.~(\ref{eq:dir-2dlam1}) is an asymptotic one and is only valid for
small $\beta$.

It is easy to see that in Eq.~(\ref{eq:dir-2dlam1}) there are only even powers of $\beta$ since the plus or the minus sign is taken with equal probability. With different probabilities, {\em e.g.}, $p(+)=0.2\,,p(-)=0.8$, we have
\begin{eqnarray}\label{eq:dir-2dlam2}
\nonumber
\lambda(\beta)=&&4\ln(2)-\ln(5)-\frac{237}{256}\beta-\frac{100171}{131072}\beta^2 \\
\nonumber
               &&+\frac{10169653}{8388608}\beta^3+\frac{8210001997}{17179869184}\beta^4 \\
\nonumber
               &&+\frac{1202188020939}{5497558138880}\beta^5-\frac{5433460372890823}{1055531162664960}\beta^6 \\             
               &&+O(\beta^7),
\end{eqnarray}
which is plotted in Fig.~\ref{f:dir-2dlam}(b) and matches well with the simulation result. Here, in the
expression Eq.~(\ref{eq:dir-2dlam2}), both even and odd powers of $\beta$ are present as the symmetry in the sign of $\beta$ is lost. To further see the trend, we take $p(+)=0.4\,,p(-)=0.6$ which gives
\begin{eqnarray}\label{eq:dir-2dlam3}
\nonumber
\lambda(\beta)=&&4\ln(2)-\ln(5)-\frac{79}{256}\beta-\frac{149819}{131072}\beta^2 \\
\nonumber
               &&+\frac{11994217}{25165824}\beta^3+\frac{11371586637}{17179869184}\beta^4 \\
\nonumber
               &&+\frac{4828258136419}{16492674416640}\beta^5-\frac{10112843356609571}{3166593487994880}\beta^6 \\
               &&+O(\beta^7),
\end{eqnarray}
which is plotted in Fig.~\ref{f:dir-2dlam}(c). From the above three plots and more plots not shown, we can see that the analytic result agrees with the simulation in a decreasing region as $p(+)$ increases which is probably due to the complex structure emerging near $\beta \sim 0.42$. Also, as $p(+)$ approaches $0.5$, the coefficients of the odd powers of $\beta$ become smaller and smaller, which vanishes at $\beta=0.5$.\\
To check the validity of the method in more complicated cases, we consider the following situation
\begin{equation}
A=\left(\begin{array}{cc}  3 & 0.3 \\
                           2 & 0.2  \end{array}\right) + \beta B_i\,,i=1,2,3
\,,\label{eq:2dp-dir1}
\end{equation}
where
\begin{equation}
B_1=\left(\begin{array}{cc}3 & 2 \\
                           6 & 5  \end{array}\right)\,,
B_2=\left(\begin{array}{cc}4 & 2 \\
                           1 & 5  \end{array}\right)\,,
B_3=\left(\begin{array}{cc}3 & 5 \\
                           1 & 2  \end{array}\right)
\,,\label{eq:2dp-b}
\end{equation}
and $p(1)=1/6\,,p(2)=1/6\,,p(3)=2/3$. Our iteration gives
\begin{eqnarray}\label{eq:2dp-lam1}
\nonumber
\lambda(\beta)=&&4\ln(2)-\ln(5)+\frac{1865}{1024}\beta-\frac{47835425}{28311552}\beta^2 \\
\nonumber
               &&+\frac{16396501375}{7247757312}\beta^3-\frac{198183253204375}{44530220924928}\beta^4 \\
\nonumber
               &&+\frac{208937303364213625}{17099304835172352}\beta^5 \\
\nonumber
               &&-\frac{1976335641588741029125}{52529986053649465344}\beta^6 \\
               &&+O(\beta^7),
\end{eqnarray}
which is plotted and compared favourably with the numerical result in Fig.~\ref{f:dir-2dlam}(d).
\subsection{Expansion for 3-d and 4-d matrices}
Our scheme is independent of the dimension of the involved matrices. Below, we show the computation on the
$3\times 3$ or $4\times 4$ matrices. Here, we give one example for each case.
\begin{figure*}
\begin{minipage}{0.48\linewidth}
\centerline{\includegraphics[width=8.5cm]{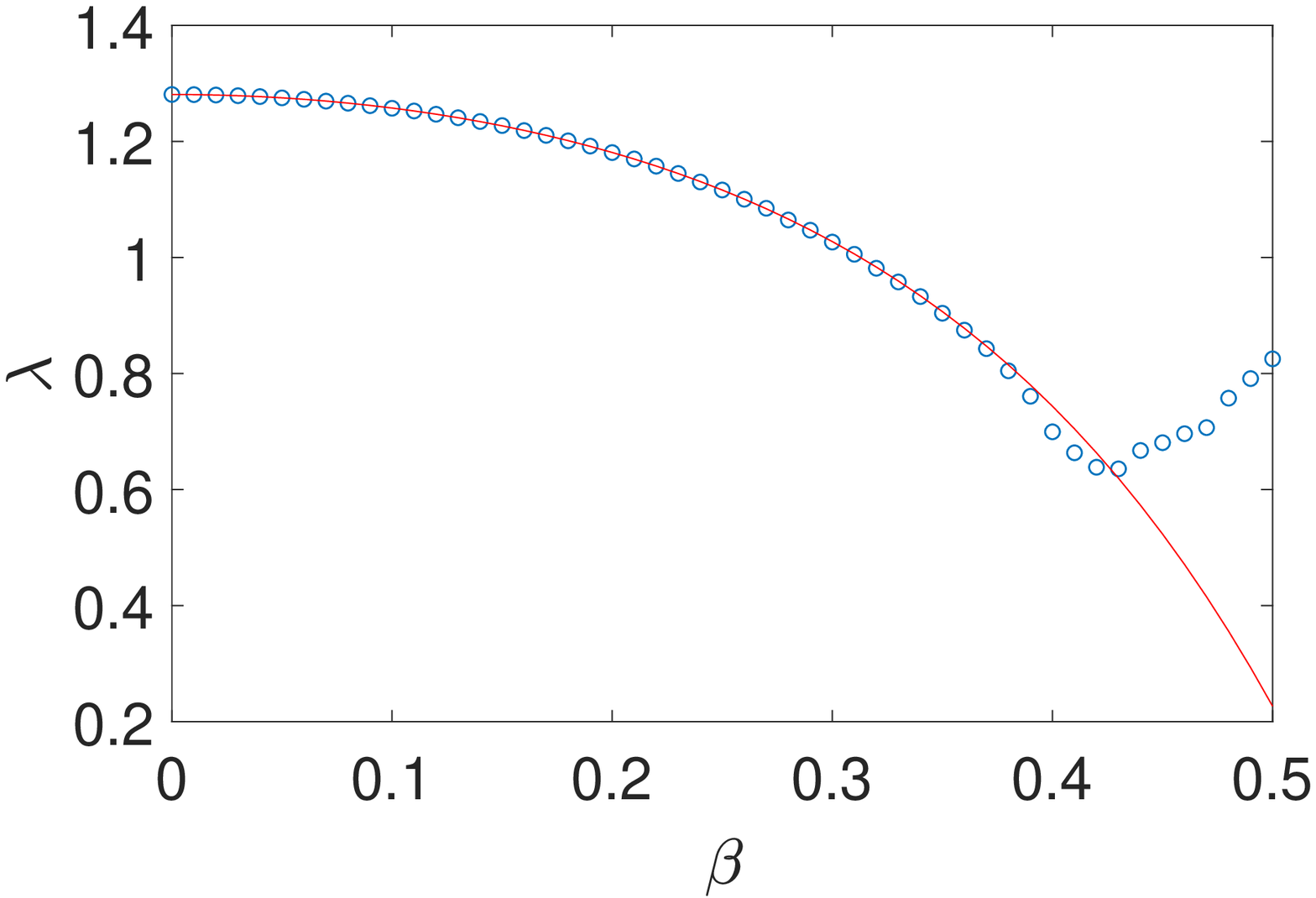}}
\centerline{(a)}
\end{minipage}
\hfill
\begin{minipage}{0.48\linewidth}
\centerline{\includegraphics[width=8.5cm]{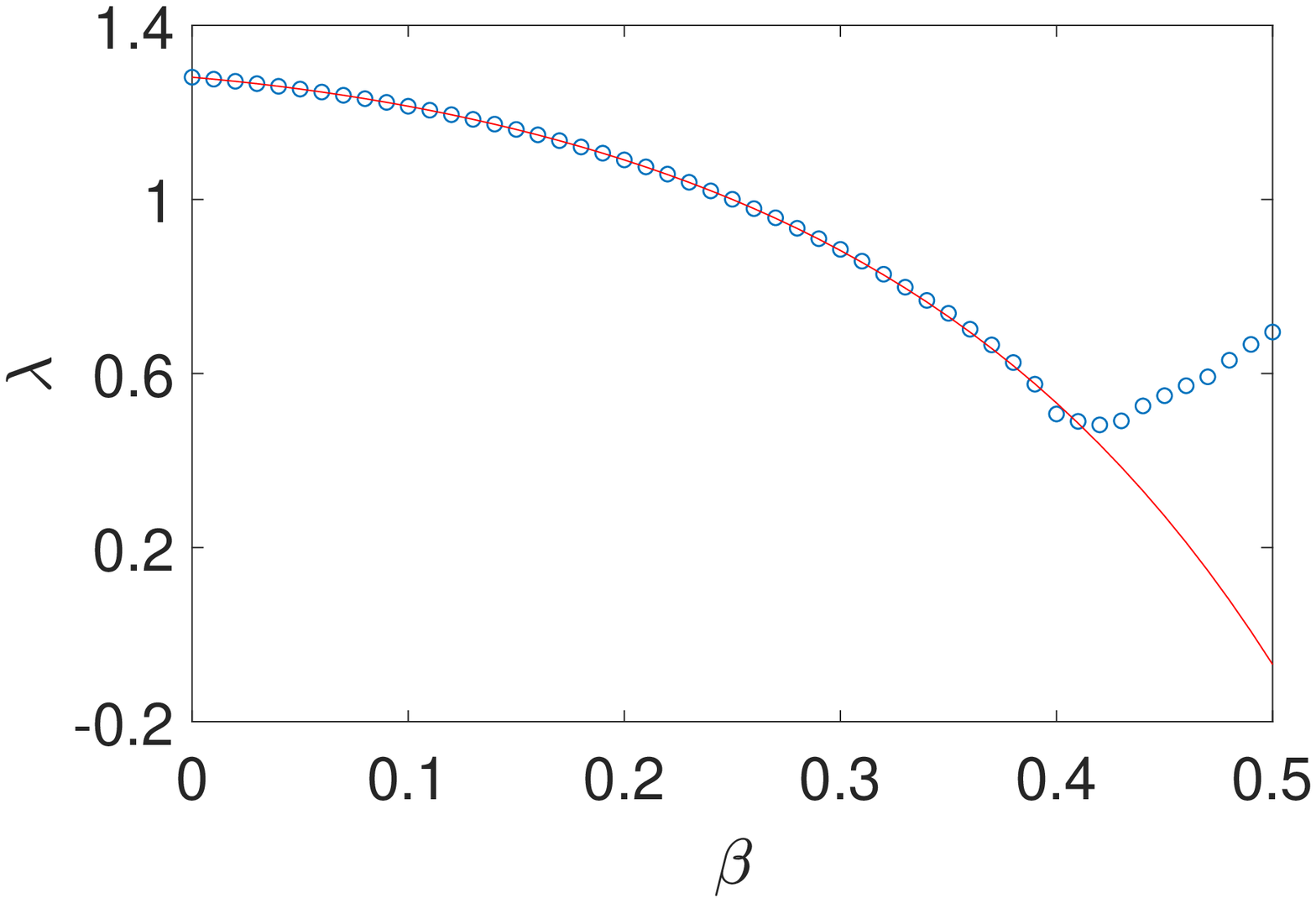}}
\centerline{(b)}
\end{minipage}
\vfill
\begin{minipage}{0.48\linewidth}
\centerline{\includegraphics[width=8.5cm]{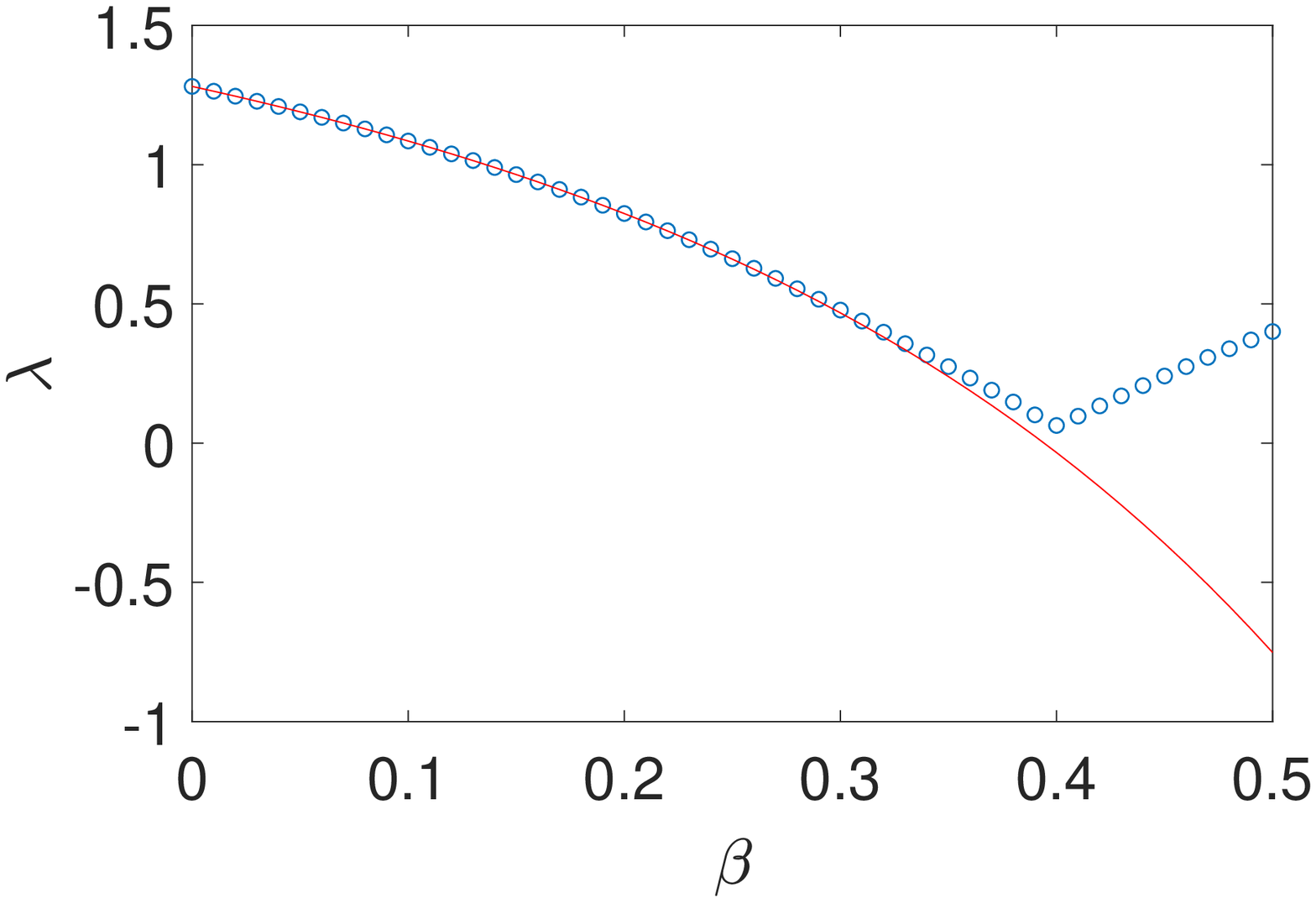}}
\centerline{(c)}
\end{minipage}
\hfill
\begin{minipage}{0.48\linewidth}
\centerline{\includegraphics[width=8.5cm]{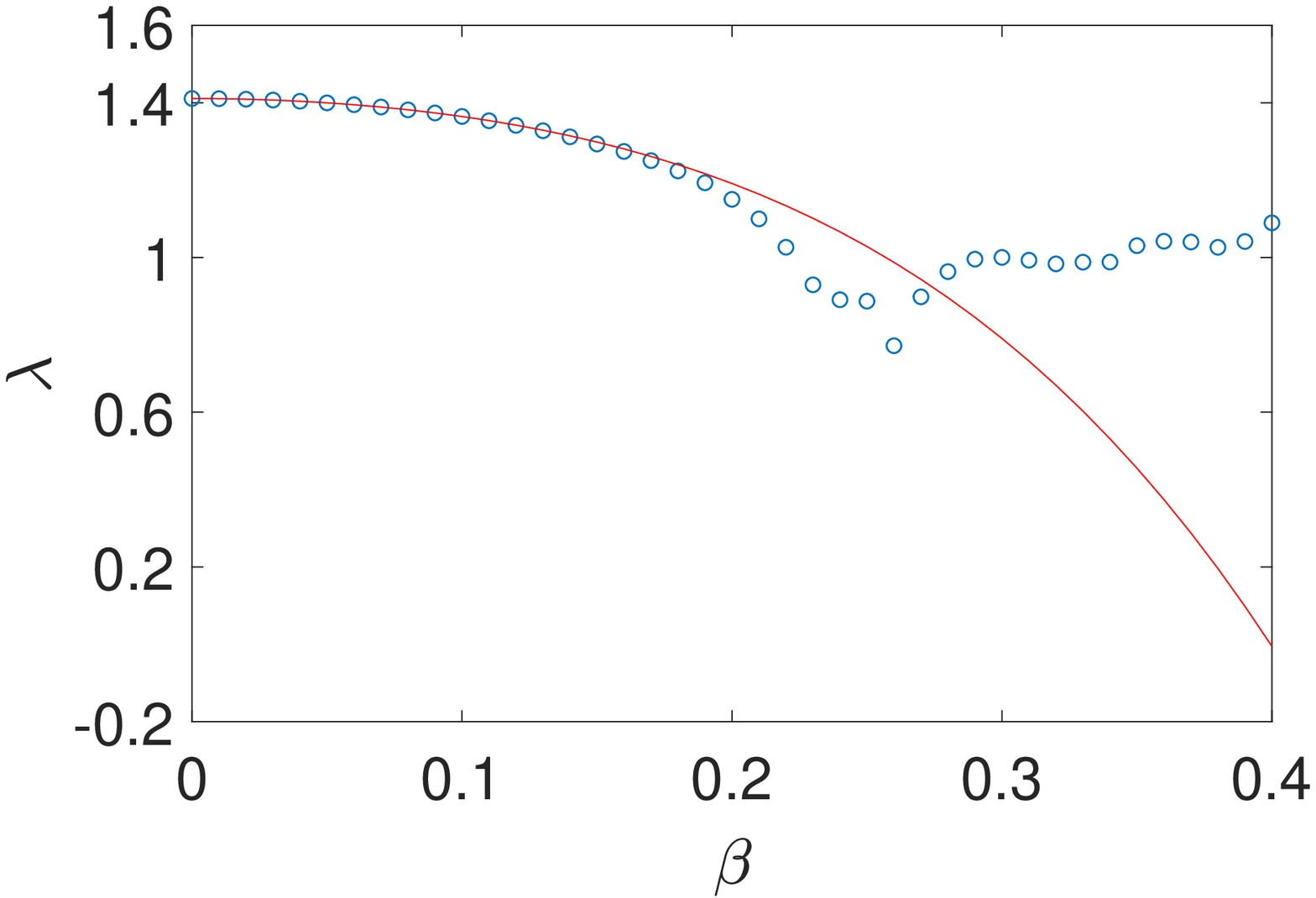}}
\centerline{(d)}
\end{minipage}
\caption{(Color online) The dependence of the MLE on
$\beta$. For  Eq.~(\ref{eq:3d-dir}) with (a)$p(+)=p(-)=0.5$; (b)$p(+)=0.4\,,p(-)=0.6$; (c)$p(+)=0.1\,,p(-)=0.9$; (d) For the system given by Eq.~(\ref{eq:4d-dir}) with $p(+)=p(-)=0.5$. The Monte carlo results are marked with (blue) circles while the analytic approximation is plotted with (red) solid lines.
} \label{f:dir-3dlam}
\end{figure*}
If we take the 3d matrices to be
\begin{equation}
A=\left(\begin{array}{ccc} 3 & 0.3 & 0.3\\
                          2 & 0.2  & 0.2\\
                          4 & 0.4 & 0.4 \end{array}\right) \pm \beta
    \left(\begin{array}{ccc}5 & 2 & 1\\
                           2 & 1  & 3\\
                            3 & 1 & 4 \end{array}\right)
\,\label{eq:3d-dir}
\end{equation}
with $p(+)=p(-)=0.5$ then the MLE is
\begin{eqnarray}\label{eq:dir-3dlam1}
\nonumber
\lambda(\beta)=&&\ln(2)+2\ln(3)-\ln(5)-\frac{1965145}{839808}\beta^2 \\
\nonumber
               &&-\frac{3740078636935}{705277476864}\beta^4-\frac{2664607365123862205}{22211625233825792}\beta^6 \\
               &&+O(\beta^8).
\end{eqnarray}
As before, the symmetry in the plus and minus sign results in even powers of $\beta$.
As shown in Fig.~\ref{f:dir-3dlam}(a), the analytic result Eq.~(\ref{eq:dir-3dlam1}) matches well
with the numerical result when $\beta<0.38$. If we take $p(+)=0.4\,,p(-)=0.6$, then
\begin{eqnarray}\label{eq:dir-3dlam2}
\nonumber
\lambda(\beta)=&&\ln(2)+2\ln(3)-\ln(5)-\frac{277}{648}\beta-\frac{363133}{155520}\beta^2 \\
\nonumber
               &&-\frac{1155711217}{2040733440}\beta^3-\frac{460024252888567}{88159684608000}\beta^4 \\
\nonumber              
               &&-\frac{903051494441389}{2380311484416000}\beta^5 \\
               &&-\frac{56063372931443174652157}{4916533371061248000000}\beta^6+O(\beta^7),
\end{eqnarray}
which is plotted in Fig.~\ref{f:dir-3dlam}(b). If we take $p(+)=0.1\,,p(-)=0.9$, then
\begin{eqnarray}\label{eq:dir-3dlam3}
\nonumber
\lambda(\beta)=&&\ln(2)+2\ln(3)-\ln(5)-\frac{277}{162}\beta-\frac{3162527}{1399680}\beta^2 \\
\nonumber
               &&-\frac{2280847319}{1020366720}\beta^3-\frac{349001178994027}{88159684608000}\beta^4 \\
\nonumber
               &&-\frac{10734200281829}{6198727824000}\beta^5 \\
               &&-\frac{47383578974881029764663}{1966613348424499200000}\beta^6+O(\beta^7),
\end{eqnarray}
which is plotted in Fig.~\ref{f:dir-3dlam}(c). From Eq.~(\ref{eq:dir-3dlam1}),(\ref{eq:dir-3dlam2}) and (\ref{eq:dir-3dlam3}), we see that the magnitude of the coefficients of the odd powers in the expansion increases with the decrease of $p(+)$, while the validity domain of the expansion increases.\\
For the 4d case, we take
\begin{equation}
A=\left(\begin{array}{cccc} 3 & 0.3 & 0.3 & 0.3\\
                          2 & 0.2  & 0.2 & 0.2\\
                          4 & 0.4 & 0.4 & 0.4 \\
                          5 & 0.5 & 0.5 & 0.5 \end{array}\right) \pm \beta
    \left(\begin{array}{cccc}5 & 2 & 1 & 3\\
                           2 & 1  & 3 & 4\\
                            3 & 1 & 1 & 5\\
                            1 & 3 & 2 & 4 \end{array}\right)
\,\label{eq:4d-dir}
\end{equation}
with $p(+)=p(-)=0.5$, the MLE is then
\begin{eqnarray}\label{eq:dir-4dlam1}
\nonumber
\lambda(\beta)=&&\ln(41)-\ln(2)-\ln(5)-\frac{11900886}{2825761}\beta^2 \\
               &&-\frac{228277459672908}{7984925229121}\beta^4+O(\beta^6),
\end{eqnarray}
which is plotted in Fig.~\ref{f:dir-3dlam}(d) together with the numerical result.
From the figure, we can see that the validity region of Eq.~(\ref{eq:dir-4dlam1})
reduces again compared to the 2d and 3d case. Also, the dependence on $\beta$ becomes quite
complex after $\beta \sim 0.2$.
\subsection{Direct expansion for continuous distribution}
\label{sect:contin}
\begin{figure}[htbp]
\begin{minipage}{0.48\linewidth}
\centerline{\includegraphics[width=8.5cm]{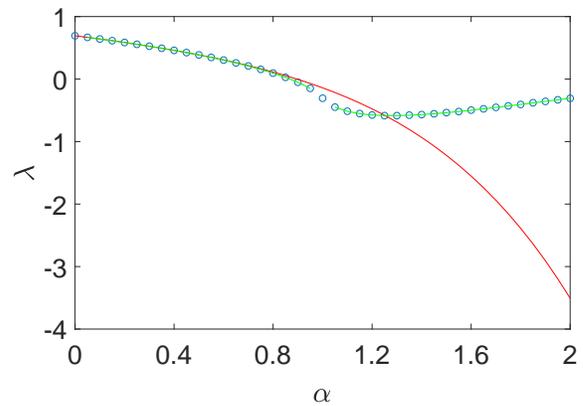}}
\end{minipage}
\caption{(Color online) The dependence of the MLE on
$\beta$ for the system described by Eq.~(\ref{eq:2d-con1}). The Monte carlo result is marked with (blue) circles while the exact one is plotted with (green) solid lines. The analytic approximation is plotted with (red) solid lines.
} \label{f:dir-2dcon}
\end{figure}

We use one example to demonstrate the application of the current technique to matrices with continuous distribution. With the specific matrices and distribution below, we are able to derive simple accurate expression of $\lambda$, with which the series expansion is compared. \\
We use the $2 \times 2$ matrices 
\begin{equation}
A=\left(\begin{array}{cc}  1 & 1 \\
                           1 & 1  \end{array}\right) + \beta
    \left(\begin{array}{cc}0 & -2 \\
                           0 & -2 \end{array}\right)
\,.\label{eq:2d-con1}
\end{equation}
The distribution of $\lambda$ is uniform in the interval $[0\,,\alpha]$ with
\begin{equation}
p(\beta)=\frac{1}{\alpha}
\,,\label{eq:dis}
\end{equation}
for given $\alpha>0$. The substitution in  Eq.~(\ref{eq:n-zi}) becomes
\begin{equation}
\tilde{z}_2=1-2\beta
\,.\label{eq:n-z2}
\end{equation}
In this particular case, after one iteration, $z_2$ in the expression $f_n$ disappears and thus $f_n$ becomes a constant. In other words, an accurate expression of $\lambda$ is generated which is written as
\begin{equation}
\lambda(\alpha)=\frac{(\alpha-1)\ln(1-\alpha)+\alpha(\ln(2)-1)}{\alpha}
\,,\label{eq:2dcon1}
\end{equation}
and its direct expansion around $\alpha=0$ is
\begin{eqnarray}\label{eq:2dcon1sym}
\nonumber
\lambda(\alpha)=&&\ln(2)-\frac{1}{2}\alpha-\frac{1}{6}\alpha^2-\frac{1}{12}\alpha^3 \\
                &&-\frac{1}{20}\alpha^4-\frac{1}{30}\alpha^5-\frac{1}{42}\alpha^6+O(\alpha^7),
\end{eqnarray}
Both of these expressions are plotted in Fig.~\ref{f:dir-2dcon} and compared with the results from the Monte Carlo simulation. It is easy to
see that the exact result matches well with the Monte carlo simulation for all $\alpha$. Nevertheless, the asymptotic expansion does not agree with the simulation when $\alpha\sim 1$. which is understandably due to the abandonment of high order terms. 
In this example, it is easy to see that the singularity at $\alpha=1$ in the logrithmic function plays a central role in the convergence of the asymptotic series. Even though the convergence is very good for small $\alpha$, near the singularity the expansion could totally go awry, which may also explain the sudden explosion of the error in the asymptotic expansion in Fig.~\ref{f:dir-2dlam} and Fig.~\ref{f:dir-3dlam}.

\section{Computation based on invariant polynomials}
\label{sec:pol}
Below, the major deterministic part of the random matrix $A$ is denoted as $A_0$, which is the matrix for $\beta=0$. Although the direct expansion technique is fast and convenient, the problem is what if the matrix $A_0$ does not have that special structure or cannot even be transformed to that structure. A direct iteration of $f_n(x)$ in Eq.~(\ref{eq:n-recur}) usually incurs exponentially more terms, which is soon out of reach. In this section, a more sophisticated expansion based on invariant polynomials is introduced to cope with this difficulty.
\subsection{A general formulation}
In the study below, the matrix $A_0$ is always considered to be diagonalizable. For simplicity, we assume that this has already been done and
\begin{eqnarray}\label{eq:a-diag}
\nonumber
&&A=\mathrm{Diag}(\lambda_1\,,\lambda_2\,,\cdots\,,\lambda_{k+1}) \\
&&\mbox{ with } |\lambda_1|>|\lambda_2|>\cdots>|\lambda_{k+1}|,
\end{eqnarray}
where $\lambda_i$'s are eigenvalues of $A$. With this simplification, the substitution in Eq.~(\ref{eq:n-T}) is greatly simplified. Especially, the deterministic part of $\tilde{z}_i$ (see Eq.~(\ref{eq:n-zi})) becomes $\lambda_i z_i/\lambda_1\,,i=2\,,3\,,\cdots\,,k+1$. If we assume that all $z_i$'s are small, all $\tilde{z}_i$'s could essentially
be approximated by polynomials of $\{z_i\}_{i=2,\cdots,k+1}$ up to a given order $m$. In fact we can do a Taylor expansion of the denominator of  Eq.~(\ref{eq:n-zi}) to the order $m$ to get these polynomials. The starting Eq.~(\ref{eq:n-start}) could also be expanded to the same order, {\em i.e.}, we may start with an $m-$th order polynomial
\begin{equation}
f_0(\mathbf{z})=\sum_{i=0}^m a_i \mathbf{z}^i
\,,\label{eq:m-poly}
\end{equation}
where the short-hand notation
\begin{equation}
\mathbf{z}^i=\Pi_{j=2}^{k+1}z_j^{i_j}\,,\mbox{ with } \sum_j i_j=i
\,,\label{eq:m-nota}
\end{equation}
and the index in the coefficient $a_i$ should be similarly understood. With the polynomial substitution and the ensuing truncation to the $m-$th order, the polynomial $f_n(\mathbf{z})$ will approach a stationary form which is invariant under the iteration, since $|\lambda_i/\lambda_1|< 1$ and the random term $B(\alpha)$ is assumed small.\\
Alternatively, we may assume that the polynomial $f_0(\mathbf{z})$ in Eq.~(\ref{eq:m-poly}) is already invariant and seek for appropriate coefficients $a_i$. Upon one iteration, these coefficients become $\tilde{a}_i$ that are linearly related to the original coefficients, {\em i.e.}
\begin{equation}
\tilde{a}_i=C_{ij} a_j
\,,\label{eq:m-linear}
\end{equation}
where $C$ is the transformation matrix that depends on the random matrices $B(\alpha)$. The invariance condition predicts that $C$ has $1$ as its dominant eigenvalue and the absolute values of all other eigenvalues are smaller than $1$. The corresponding left eigenvector $L_i$ and right eigenvector $R_i$ could be obtained by solving the linear equation
\begin{equation}
(C_{ij}-\delta_{ij})R_j=0\,,\mbox{ and } (C_{ij}-\delta_{ij})L_i=0
\,,\label{eq:m-eigeq}
\end{equation}
where $\delta_{ij}$ is the Kronecker delta function.\\
The starting function Eq.~(\ref{eq:n-start}) could of course be expanded into a polynomial form with the coefficient vector $b_i$. The final stationary coefficient vector $\bar{b}_i$ can be determined by the three vectors $b_i\,,L_i\,,R_i$ as
\begin{equation}
\bar{b}_i=\frac{\sum_j L_j b_j}{\sum_j L_j R_j}R_i
\,,\label{eq:m-bbar}
\end{equation}
since all other vectors decreases exponentially to zero upon iteration. 
From Eq.~(\ref{eq:n-lam}), the constant term $\bar{b}_0$ in the invariant polynomial gives the MLE. In fact, it is not hard to see that the right eigenvector could be taken as $R_i=\delta_{i0}$, corresponding to a zeroth order polynomial, which is obviously invariant under the iteration. Therefore,
\begin{equation}
\lambda=\frac{\sum_j L_j b_j}{\sum_j L_j R_j}
\,.\label{eq:n-lamLR}
\end{equation}
Below, we will use several examples to demonstrate application of this new scheme.
\subsection{Two examples}
\begin{figure*}
\begin{minipage}{0.48\linewidth}
\centerline{\includegraphics[width=8.5cm]{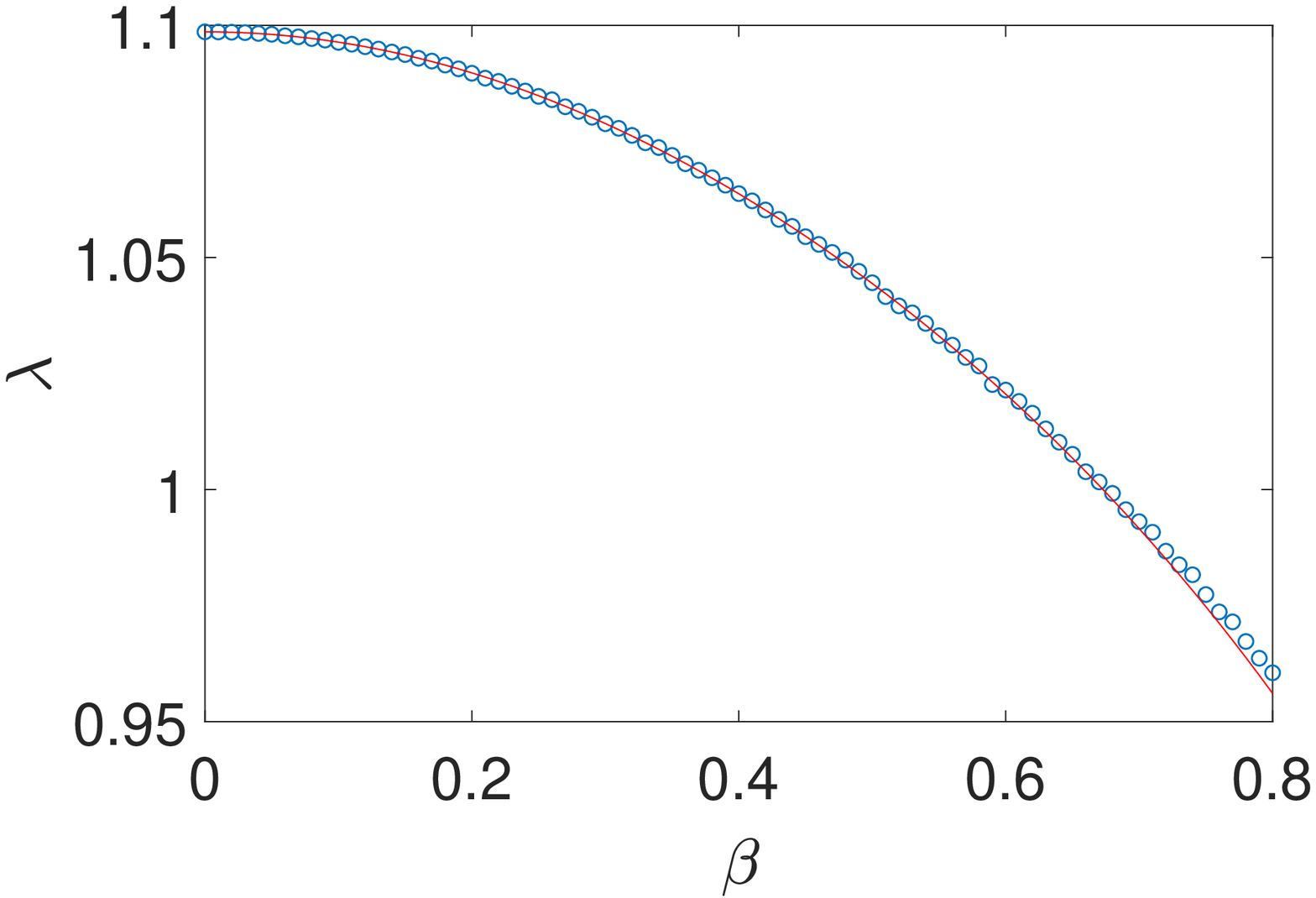}}
\centerline{(a)}
\end{minipage}
\hfill
\begin{minipage}{0.48\linewidth}
\centerline{\includegraphics[width=8.5cm]{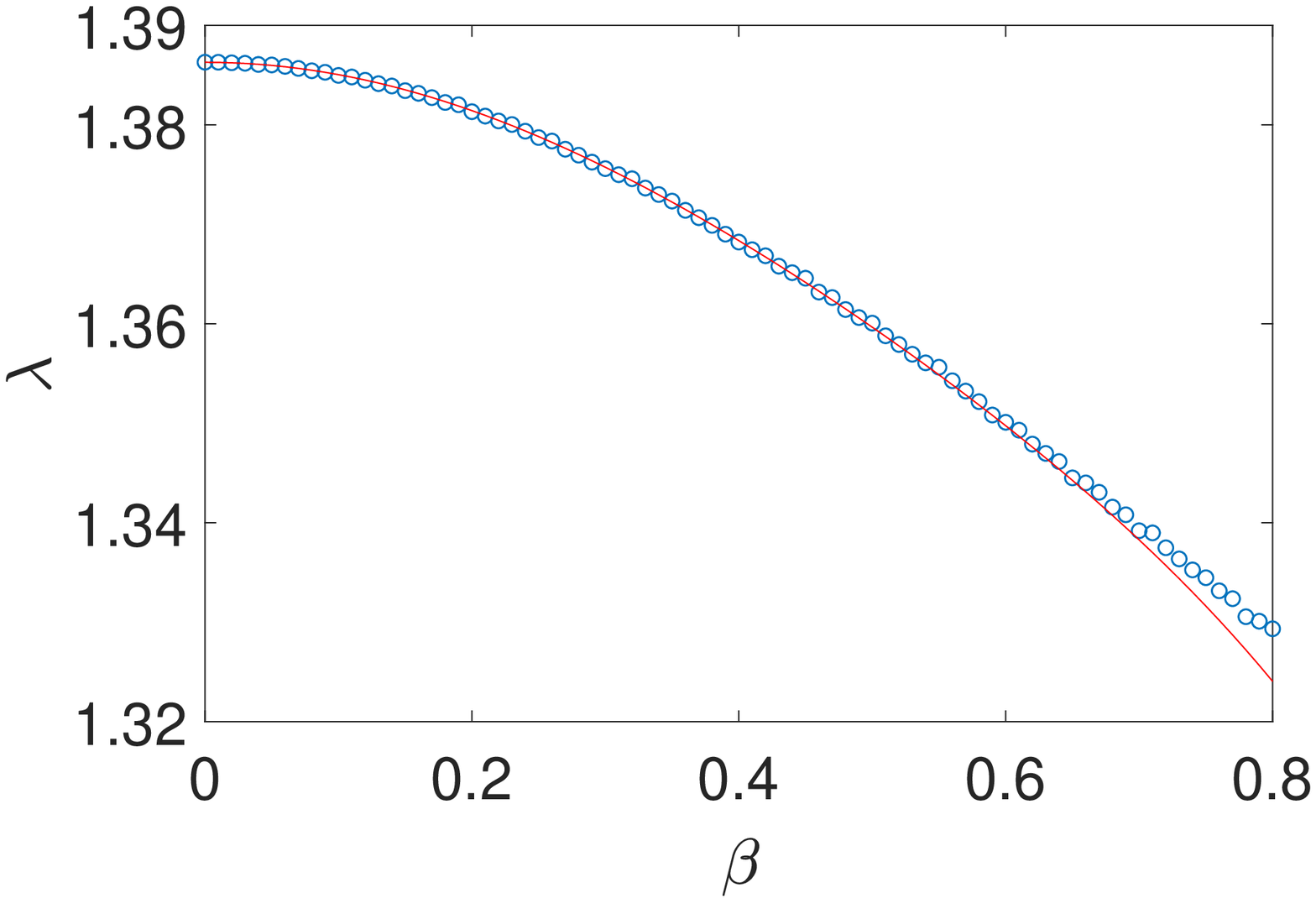}}
\centerline{(b)}
\end{minipage}
\caption{(Color online) Comparison of the expansion result with the numerical computation for the systems givey by (a) Eq.~(\ref{eq:pol-2d}), (b)Eq.~(\ref{eq:3d-pol}). The Monte carlo results are marked with (blue) circles while the analytic approximation is plotted with (red) solid lines.
}
\label{f:pol-2dlam}
\end{figure*}
Here, we give examples for the application of the above formulation. For simplicity, we assume that when
$\beta=0$, the major part of the matrix is already in the diagonal form. For the two dimensional case, we take
\begin{equation}
A_n=\left(\begin{array}{cc}3 & 0 \\
                           0 & 2  \end{array}\right) \pm \beta
    \left(\begin{array}{cc}2 & 0.5 \\
                           1.2 & 3.0  \end{array}\right)
\,,\label{eq:pol-2d}
\end{equation}
with $p(+)=p(-)=0.5$.
A direct application of the above scheme gives the growth rate of the system
\begin{equation}
\lambda(\beta)=\ln(3)-\frac{2}{9}\beta^2+\frac{719}{20250}\beta^4-\frac{774479}{13668750}\beta^6+O(\beta^8)
\,,\label{eq:pol-2dlam}
\end{equation}
where only even powers of $\beta$ are present due to the above mentioned symmetry. A comparison of Eq.~(\ref{eq:pol-2dlam}) with the numerical computation is shown in Fig.~\ref{f:pol-2dlam}(a).
It is easy to see that the two results agree very well in the whole computational region.
In the case of $3\times 3$ matrices, we use
\begin{equation}
A=\left(\begin{array}{ccc} 4 & 0 & 0\\
                          0 & 2  & 0\\
                          0 & 0 & 3 \end{array}\right) \pm \beta
    \left(\begin{array}{ccc}2 & 1.2 & 0.6\\
                           0.5 & 0.7  & 2\\
                            3 & 1 & 0.4 \end{array}\right)
\,\label{eq:3d-pol}
\end{equation}
with $p(+)=p(-)=0.5$, then the maximum Lyapunov exponent is
\begin{eqnarray}\label{eq:pol-3dlam}
\nonumber
\lambda(\beta)=&&2\ln(2)-\frac{1}{8}\beta^2+\frac{926378511}{13421772800}\beta^4 \\
               &&-\frac{1771471938223}{64424509440000}\beta^6+O(\beta^8),
\end{eqnarray}
which is plotted in Fig.~\ref{f:pol-2dlam}(b) together with the numerical profile. The two agrees
quite well until $\beta \sim 0.7$, which still confirms the
validity of our new scheme based on invariant polynomials.
\section{Application to the Ising model with random on-site fields}
\label{sec:isi}
\begin{figure}
\begin{minipage}{0.48\linewidth}
\centerline{\includegraphics[width=8.5cm]{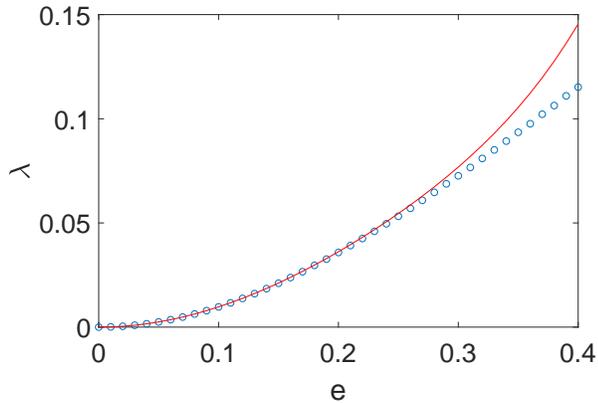}}
\end{minipage}
\caption{(Color online) A comparison of the expansion with the numerical simulation for Eq.~(\ref{eq:is-lambda2}). The Monte carlo results are marked with (blue) circles while the analytic approximation is plotted with (red) solid lines.
}
\label{f:isi-lam}
\end{figure}

In this section, we apply our method to the one-dimension Ising model with random on-site field. 
In this model, $N$ spins lay side by side along a one-dimensional lattice, each of which has two possible states: the spin up state ($\sigma_i=1$), or the down state ($\sigma_i=-1$).
Following the terminology from~\cite{cris93prod}, we may write down the Hamiltonian of the system
\begin{equation}
H=-J\sum_{i=1}^{N}\sigma_i\sigma_{i+1}-\sum_{i=1}^{N} h_i\sigma_i
\,,\label{eq:is-H}
\end{equation}
where $J$ is the coupling constant between nearest-neighbor spins and $h_i$ is the external magnetic field on each site, being assumed to be random. The first term in Eq.~(\ref{eq:is-H}) describes the interaction of neighboring spins and the second term takes into account the energy in the external field. The partition function is
\begin{equation}
Z_N=\sum_{\{\sigma\}} \exp(\beta J \sum_i \sigma_i \sigma_{i+1}+ \beta \sum_i h_i \sigma_i)
\,.\label{eq:is-ZN}
\end{equation}
Introducing the transfer matrix
\begin{equation}
L_i(\sigma_i,\sigma_{i+1})=\exp(\beta J \sigma_i \sigma_{i+1}+\beta h_i\sigma_i)
\,,\label{eq:is-ma1}
\end{equation}
or more explicitly
\begin{eqnarray}\label{eq:is-ma2}
\nonumber
L_i&&=\left(\begin{array}{cc} L_i(1,1) & L_i(1,-1) \\
                            L_i(-1,1) & L_i(-1,-1)  \end{array}\right) \\
   &&=\left(\begin{array}{cc} e^{\beta(J+h_i)} & e^{\beta(-J+h_i)} \\
                            e^{\beta(-J-h_i)} & e^{\beta(J-h_i)} \end{array}\right),
\end{eqnarray}
the free energy per spin could be written as 
\begin{equation}
F=-\lim_{N\to\infty}\frac{1}{\beta N}\ln Z_N=-\lim_{N\to\infty}\frac{1}{\beta N}\ln[Tr(\prod_{i=1}^NL_i)]
\,.\label{eq:is-F1}
\end{equation}
The equations above show the general procedure of computing the free energy of the 1-d Ising model using the transfer matrix approach. An exact analytic expression may be derived if the external field is constant over the sites. In the presence of local randomness, the external field $h_i$ is a quenched random variable, differing from site to site, for which it is hard to get an exact solution. When the interaction is strong, an analytic expansion is easily derived with our method. First, we reshape Eq.~(\ref{eq:is-ma2}) as follows
\begin{eqnarray}\label{eq:is-ma2re}
\nonumber
L_i&&=e^{\beta(J+h_i)}\left(\begin{array}{cc} 1 & e^{-2\beta J} \\
                                          e^{-2\beta(J+h_i)} & e^{-2\beta h_i} \end{array}\right) \\
   &&=e^{\beta(J+h_i)} \left(\begin{array}{cc} 1 & \epsilon \\
                                            \epsilon x_i & x_i \end{array}\right),
\end{eqnarray}
with $x_i=e^{-2\beta h_i}$ and $\epsilon=e^{-2 \beta J}$, being small and to be used as the expansion parameter. When the reshaped $L_i$ is used in Eq.~(\ref{eq:is-F1}), the free energy could be written as~\cite{cris93prod}
\begin{equation}
F=-J-\overline{h}-\frac{1}{\beta}\lim_{N\to\infty}\frac{1}{N}\ln[Tr(\prod_{i=1}^N\left(\begin{array}{cc} 1 & \epsilon \\
                                            \epsilon x_i & x_i \end{array}\right))]
\,.\label{eq:is-F2}
\end{equation}
$\overline{h}$is the average of $h_i$. Once we give the distribution of $\epsilon$, the first and the second term in Eq.~(\ref{eq:is-F2}) is fixed. What is left to do is to calculate the third term's analytical result. Since the major part of the third term is the standard form of the Lyapunov Exponent of ramdon matrices, we may apply our method into it.
\begin{equation}
\lambda=\lim_{N\to\infty}\frac{1}{N}\ln[Tr(\prod_{i=1}^N\left(\begin{array}{cc} 1 & \epsilon \\
                                            \epsilon x_i & x_i \end{array}\right))]
\,,\label{eq:is-lambda1}
\end{equation}
In ~\cite{cris93prod}, $\overline{x}$ which is the average of $x_i$ should be smaller than $1$, so we treat $x_i$ as uniformly distribution in $(0,1)$, which not only can satisfy the above condition and but also is a common assumption when it comes to random field. Then the second term we mentioned in Eq.~(\ref{eq:is-F2}) can be generated.
\begin{equation}
P(x)=1
\,,\label{eq:is-P}
\end{equation}
\begin{equation}
\tilde{z}_2=\frac{\epsilon+x z_2}{1+\epsilon x z_2}
\,,\label{eq:is-z2}
\end{equation}
and
\begin{equation}
f_0(z_2)=\int_{0}^{1}P(x)\ln(1+\epsilon x z_2)\,dx
\,.\label{eq:is-f0}
\end{equation}
After a few iterations, the result is 
\begin{equation}
\lambda(\epsilon)=\epsilon^2-\frac{269}{90}\epsilon^4+\frac{392309}{25920}\epsilon^6+O(\epsilon^8)
\,.\label{eq:is-lambda2}
\end{equation}
In ~\cite{cris93prod}, there is a relation between $\overline{x}$ and the second order term's coefficient of $\lambda$,
\begin{equation}
S_2=\frac{\overline{x}}{1-\overline{x}}
\,,\label{eq:is-rela}
\end{equation}
where $S_2$ is the coefficient of the second order term. With the current distribution of $x$, the coefficient is $1$ which agrees with our result. In Fig.~\ref{f:isi-lam}, we see 
that analytic result matches well with the numerical result when $\epsilon<0.3$. The free energy of the system reads

\begin{eqnarray}\label{eq:is-F3}
\nonumber
F=&&-\frac{1}{2}-J-\frac{e^{-4 \beta J}}{\beta}-\frac{269e^{-8 \beta J}}{90 \beta} \\
  &&+\frac{392309e^{-12 \beta J}}{25920 \beta}+O(e^{-16 \beta J}).
\end{eqnarray}
The taylor expasion we use during the process assumes $\epsilon$ is small which means the Eq.~(\ref{eq:is-F3}) matches the real situation only if $J$ is large enough. 
\section{Discussion}
\label{sec:dis}
In this paper, we generalize the generating function approach developed in \cite{lan11,lan13} for dealing with stochastic sequences to the treatment of random matrix products. After a general formalism is introduced, two analytic schemes are designed for the computation of the MLE: a direct iteration scheme and a scheme based on invariant polynomials. Several examples are given to illustrate the usage of the methods. As expected, when the randomness is small ($\beta$ small), the analytical results agree very well with the numerical ones. However, with the increase of randomness, the agreement becomes worse and higher order terms may chip in playing a role. In many cases, the increase of randomness brings in complex dependence of the MLE on $\beta$ which can not be captured by the analytic expansion~\cite{tre01comp,emb99tre}. When the current technique is used in the study of an Ising model with random onsite fields, the free energy is conveniently obtained with an analytic approximation in case of a strong coupling between neighboring spins.  \\
The direct iteration scheme is very easy to implement but requires a special form of the major part of the random matrix
which may not be present in an application. The second scheme based on invariant polynomials is very general but requires solution of a linear equation. Also, we assume that the major part of the random matrix has one
dominant eigenvalue to guarantee a good convergence of the expansion. If there are multiple eigenvalues that have the same magnitude, the convergence of
the polynomial coefficients is not guaranteed and we have to modify the method.
On the other hand, as implemented in \cite{lan11,lan13}, with the generating function formalism, a numerical scheme could be easily designed
for the computation of the growth rate for any value of $\beta$, which is at least good for small matrices. For large matrices, however, the direct numerical computation seems unmanageable. Whether it is possible to extend either of the current two analytic schemes to highly efficient numerical tools
is an interesting challenge.\\
In the examples given above, matrices with dimensions only up to $4$ are used although the formulation is valid for general random matrices. Of course, the computation could become quite cumbersome in high-dimensional cases. Also, only two or three matrices are used for the random multiplication in the above examples. It is not hard to see that the method could be easily applied to cases with more random matrices. Even for a continuous distribution of matrices, the current scheme is still
applicable~\cite{lan11}, as demonstrated in the example in Section \ref{sect:contin}.\\
In all the above computation, only the MLE is computed. How to generalize the current formulation to the evaluation of other Lyapunov exponents is still under investigation. Also, in all the calculations, we
assume that the MLE exists which measures the exponential increase of the length of a typical vector following the dynamics determined by random matrix multiplication. However, in certain critical cases, this growth may not be exponential but follow a power law~\cite{tre01comp}. How to adjust the above computation to this case is still an open problem.
\section*{Acknowledgements}
 This work was supported by the National Natural Science Foundation of China under Grants No. 11775035 and No. 11375093, and also by the Fundamental Research Funds for the Central Universities with contract number 2019XD-A10.

\end{document}